\documentclass{PoS}

\title{Testing New TeV-scale Seesaw Mediators\\
at the LHC}

\ShortTitle{TeV-scale Seesaw Mediators}

\author{\speaker{Ivica Picek}$^{,1}$ and Branimir Radov\v{c}i\'c$^{1,}$\thanks{This work is supported by the Croatian Ministry  of Science, Education and Sports under Contract No. 119-0982930-1016.}\\
         $^1$Department of Physics, University of Zagreb, Croatia\\
        E-mail: \email{picek@phy.hr}, ~\email{bradov@phy.hr}}

\abstract{We propose TeV-scale Dirac fermions producing Majorana masses of the known neutrinos via 
tree-level seesaw, different from standard type I and III seesaw. The employed weak five-plet with
nonzero hypercharge leads to new seesaw formula $m_{\nu} \sim v^6/M^5$ and to empirical
masses $m_{\nu} \sim 10^{-1}$ eV for $M \sim$ TeV new states. For a limited range of the parameter 
space, where $M \leq$ a few 100 GeV, the proposed mechanism is testable at the LHC
via characteristic decays of Dirac type heavy leptons, produced by a Drell-Yan fusion.}

\FullConference{35th International Conference of High Energy Physics - ICHEP2010,\\
		July 22-28, 2010\\
		Paris France}

\begin{document}


Our model \cite{PiR09} is  based on the standard model group (SMG) $SU(3)_C
\times SU(2)_L \times U(1)_Y$ content extended by vectorlike Dirac 5-plets of leptons,
$\Sigma_{L,R} = (\Sigma^{+++}, \Sigma^{++}, \Sigma^+, \Sigma^0, \Sigma^-)_{L,R}\ \sim\ (1,5,2)$, which form a Dirac mass term
${\cal L}_{mass} =  -\,  M_{\Sigma} \, \overline \Sigma_L \Sigma_R +  H.c.$. They, in conjunction with additional scalar four-plets $\Phi_1=(\Phi^{0}_1,
\Phi^{-}_1, \Phi^{--}_1, \Phi^{---}_1)$ and $\Phi_2=(\Phi^{+}_2
\Phi^{0}_2, \Phi^{-}_2, \Phi^{--}_2)$ $\sim$ $(1,4,-3)$ and $(1,4,-1)$ respectively, build the gauge invariant Yukawa terms
\begin{equation}
{\cal L}_{Y} = Y_1 \, \overline l_{L} \Sigma_R
\, \Phi_1  +Y_2 \, \overline \Sigma_L (l_{L})^c \, \Phi^*_2   + \mathrm{H.c.}\ .
\label{Yuk}
\end{equation}
Due to the induced {\em vevs} $v_{\Phi_1}$ and $v_{\Phi_2}$ of new scalar fields, the light neutrino acquires a mass
\begin{equation}\label{dim9}
    m_{\nu} \sim \frac {Y_1 Y_2 v_{\Phi_1} v_{\Phi_2}} {M_{\Sigma}} 
= \frac {Y_1 Y_2\ \lambda_1\lambda_2\ v^6} {M_{\Sigma}\ \mu^2_{\Phi_1}\ \mu^2_{\Phi_2}} \sim v^6/M^5 \,,
\end{equation}
corresponding to seesaw mechanism generated by dimension nine operator shown in the figure.

\hspace{-3cm}
\vspace{0.3cm}

\includegraphics[scale=0.5]{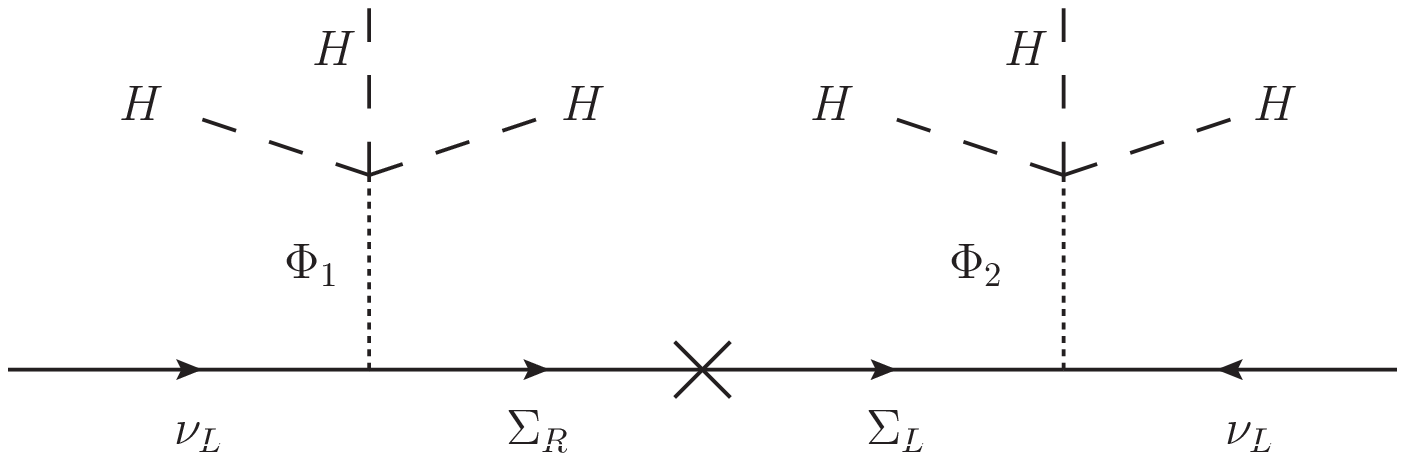}

\vspace{-3.1cm}

\begin{center}
\hspace{7cm}
\begin{tabular}{|c|c|} \hline
\small{Produced 5-plet pair}      & \small{$\sigma$/fb}    \\ \hline
 \small{$\Sigma^{+++} \overline{\Sigma^{++}}$}    &  \small{180}  \\
\small{$\Sigma^{++} \overline{\Sigma^{+}}$}    &  \small{270}  \\
   \small{$\Sigma^{+} \overline{\Sigma^{0}}$}    &  \small{270}  \\
  \small{$\Sigma^{+++} \overline{\Sigma^{+++}}$}    &  \small{230}  \\
  \small{$\Sigma^{-} \overline{\Sigma^{-}}$}    &  \small{210}  \\ \hline
\end{tabular}\\
\end{center}
\vspace{0.3cm}
The new physics scale $\Lambda_{NP}$ of our model can be estimated from eq.~(\ref{dim9}) for some reasonable values of Yukawa and 
quartic coupling strengths $\lambda_1$ and $\lambda_2$, say $Y_1\sim Y_2\sim \lambda_1\sim \lambda_2\sim10^{-2}$. Assuming degenerate all high scale mass parameters, we obtain $\Lambda_{NP}\simeq580$ GeV.

Accordingly, the 5-plet seesaw mediators can be produced by $W^{\pm}$, $Z^0$ and $\gamma$ Drell-Yan processes at the LHC. The associated production of the pairs $(\Sigma^{+++},\overline{\Sigma^{++}})$, $(\Sigma^{++},\overline{\Sigma^{+}})$, $(\Sigma^{+},\overline{\Sigma^{0}})$, $(\Sigma^{0},\overline{\Sigma^{-}})$ via a charged current would be a crucial test of the five-plet nature of new leptons. Some highest production rates are displayed in the enclosed table for $\sqrt{s}=14$ TeV (to be reduced
by roughly a factor of 5 for the present LHC energy of $\sqrt{s}=7$ TeV).
They are higher than for type III seesaw mediators, because of both additional and enhanced 
couplings to the gauge bosons. 

The produced five-plet states might be recognized via characteristic decays addressed in detail elsewhere \cite{KPiR}.
Since these states  are characterized by small mass splitting within a multiplet \cite{Cirelli:2005uq}, their cascade decays are suppressed. A distinguished triply charged fermion which doesn't mix with SM leptons has a characteristic decay,
$ \Sigma^{+++} \to W^+ W^+ l^+$ .
The singly charged and neutral states mix with SM particles, so that the produced pairs lead to
\begin{equation}\label{jets}
\Sigma^{+} \overline{\Sigma^{0}} \to l^+ Z l^+ W^- \to l^+ l^+ + 4 jets\ \ ,\ \ 
\Sigma^{0} \overline{\Sigma^{0}}  \to l^\pm W^\mp l^\pm W^\mp \to l^\pm l^\pm + 4 jets\ ,
\end{equation}
the lepton number violating decays with same sign dileptons and the jets as an appealing signature.


\begin{thebibliography}{99}

\bibitem{PiR09}
I. Picek and B. Radov\v{c}i\'c,
Phys. Lett. {\bf B 687} 338 (2010)
[{\tt 0911.1374 [hep-ph]}].

\bibitem{KPiR}
K. Kumeri\v{c}ki, I. Picek and B. Radov\v{c}i\'c,
[in preparation].

\bibitem{Cirelli:2005uq}
M. Cirelli, N. Fornengo and A. Strumia,
Nucl. Phys.
{\bf B753} (2006) 178
[{\tt hep-ph/0512090}].



\end{thebibliography}
\end{document}